\documentclass[9pt,twocolumn,twoside]{optica}
\setboolean{shortarticle}{false}
\setboolean{minireview}{false}

\newcommand{\aopd}{\hat{a}^\dagger}
\newcommand{\aop}{\hat{a}}

\newcommand{\bopd}{\hat{b}^\dagger}
\newcommand{\bop}{\hat{b}}
\newcommand{\copd}{\hat{c}^\dagger}
\newcommand{\cop}{\hat{c}}

\newcommand{\aopdp}{\hat{a'}^\dagger}
\newcommand{\aopp}{\hat{a'}}
\newcommand{\bopdp}{\hat{b'}^\dagger}
\newcommand{\bopp}{\hat{b'}}

\usepackage{braket}

\title{Cryogenic microwave-to-optical conversion using a triply-resonant lithium niobate on sapphire transducer}
\author[1,*]{Timothy P. McKenna}
\author[1,*]{Jeremy D. Witmer}
\author[1]{Rishi N. Patel}
\author[1]{Wentao Jiang}
\author[1]{Rapha\"el Van Laer}
\author[1]{Patricio Arrangoiz-Arriola}
\author[1]{E. Alex Wollack}
\author[1]{Jason F. Herrmann}
\author[1]{Amir H. Safavi-Naeini}

\affil[1]{Ginzton Laboratory, Stanford University, 348 Via Pueblo Mall, Stanford, CA 94305, USA}
\affil[*]{These authors contributed equally. Corresponding authors: tmckenna@stanford.edu, jwitmer@stanford.edu, safavi@stanford.edu}

\begin{abstract}
Quantum networks are likely to have a profound impact on the way we compute and communicate in the future.  In order to wire together superconducting quantum processors over kilometer-scale distances, we need transducers that can generate entanglement between the microwave and optical domains with high fidelity. We present an integrated electro-optic transducer that combines low-loss lithium niobate photonics with superconducting microwave resonators on a sapphire substrate.  Our triply-resonant device operates in a dilution refrigerator and converts microwave photons to optical photons with an on-chip  efficiency of $6.6\times 10^{-6}$ and a conversion bandwidth of 20 MHz. We discuss design trade-offs in this device, including strategies to manage acoustic loss, and outline ways to increase the conversion efficiency in the future.  

\end{abstract}

\setboolean{displaycopyright}{true}

\begin{document}

\maketitle

\section{Introduction}

\subsection{Background}
As quantum technologies become more developed and the need to process and exchange quantum information grows, we will need to find ways to expand the internet to include quantum links \cite{Kimble2008a}.  The advantages of a future quantum internet include distributed quantum computation, provably secure communication, and long baseline quantum sensing \cite{Wehner2018,Pant2019,Gottesman2012}.  Already significant progress has been made towards realizing quantum networks, including terrestrial and satellite-based demonstrations of distant entanglement generation \cite{Ursin2007, Yin2017}, commercially available quantum key distribution \cite{Scarani2009}, and loophole-free Bell tests \cite{Hensen2015}.  

Quantum systems with optical frequency transitions, such as defect centers in diamond \cite{Prawer2014}, quantum dots \cite{Kloeffel2013}, and trapped ions \cite{Bruzewicz2019}, naturally interact with visible or infrared light, so connecting these systems with optical links is comparatively straightforward.  In contrast, superconducting qubits \cite{Devoret2013a,Wendin2017}, a leading technology for near-term NISQ era quantum information processing \cite{Preskill2018}, operate at microwave frequencies in the few gigahertz range and are not readily addressable with light.  There has been a tremendous effort to create entanglement between increasingly distant superconducting qubits using microwave channels \cite{Narla2016,Axline2018,Walter2018,Campagne-Ibarcq2018}, culminating in a recent demonstration by Magnard et al. of entangled qubits in two dilution refrigerators separated by five meters \cite{Magnard2020}.  However, for links at microwave frequencies to operate with an acceptably low probability of adding thermal noise photons, the physical links must be cooled to cryogenic temperatures \cite{Xiang2017,Vermersch2017}, likely making them impractical for links longer than a few tens of meters.  On the other hand, optical fiber links are inherently free from thermal noise at room temperature due to their high carrier frequency and low loss, and optical fiber can be cheaply deployed over distances of many kilometers.  Millimeter wave (60 to 300 GHz) interconnects may prove useful for intermediate distances (eg. data centers or intra-city connections) \cite{Pechal2017}, but as in today's internet, for many applications, long-haul optical links are required.

The benefits of networking superconducting qubits with optical links has spurred a great deal of research on coherent conversion between optical and microwave photons \cite{Lauk2020,Lambert2020}.  There have been a variety of approaches including electro-optomechanics \cite{Andrews2014a,Bagci2014,Forsch2020}, piezo-optomechanics \cite{ Vainsencher2016,Balram2015b,Jiang2020,Shao2019,Mirhosseini2020},  direct electro-optic coupling \cite{Rueda2016,Fan2018,Witmer2020}, magnons in yttrium iron garnet \cite{Hisatomi2016}, Rydberg atoms \cite{Vogt2019,Han2018}, and rare-earth doped crystals \cite{Bartholomew2019}.  To date the highest conversion efficiencies have been achieved using the electro-optomechanical approach \cite{Higginbotham2018}; however, the approach demonstrated here using the direct electro-optic (EO) effect (aka. Pockels effect) has the advantages that it can be microfabricated on a chip, has high-conversion bandwidth and operates at a wavelength which is readily voltage-tunable.

\begin{figure} [t!]
\centering
\includegraphics[width=\columnwidth]{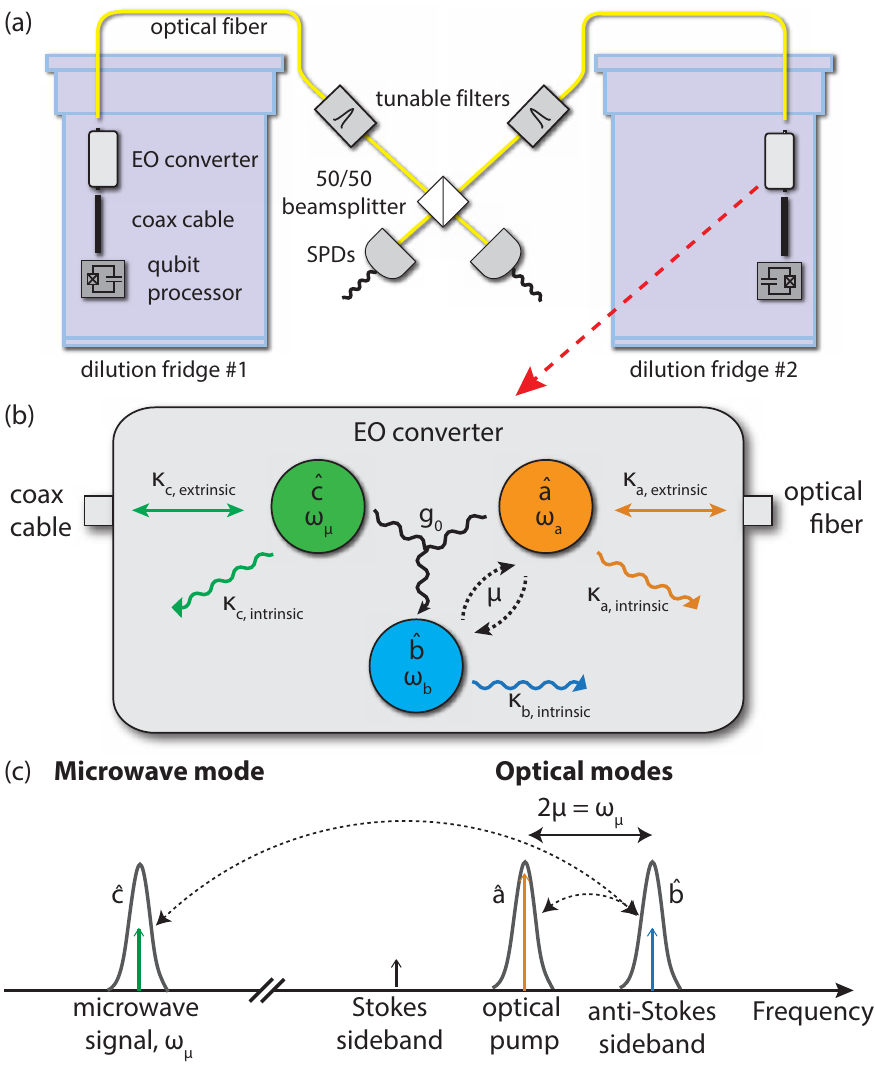}
\caption{(a)  Schematic showing a two-node quantum network connected with EO converters.  The EO converter produces entangled microwave and optical photons.  When a photon is detected by one of the single-photon detectors (SPDs), the superconducting qubits are projected into an entangled state. The tunable filters are required to remove the pump light from the EO converters.  (b) Schematic illustration of the coupled microwave and optical modes.  (c) Diagram of the EO conversion process in the case of red-side pump detuning, with the optical modes tuned to be fully hybridized.  Microwave photons are up-converted into the anti-Stokes sideband.  }
\label{fig:theory}
\end{figure}

Schematically, there are two primary approaches to use an EO converter to entangle distant qubits.  The first is using a "pitch and catch" protocol, which was first proposed in the context of cavity QED \cite{Cirac1997} and has subsequently been used for superconducting qubits \cite{Axline2018,Campagne-Ibarcq2018,Walter2018}. Here, the EO converter acts as an impedance matching element to coherently convert microwave photons to optical photons and vice versa \cite{Tsang2010,Tsang2011,Safavi-Naeini2011a}. However, this scheme has the disadvantage that it requires the converter to have near-unity conversion efficiency to create a high fidelity entangled state.  On the other hand, heralded protocols that use single-photon detection, such as the one proposed for cavity QED in \cite{Duan2001} and for microwave-to-optical converters in \cite{Zhong2020} and \cite{Rueda2019}, will be more robust in the presence of imperfect components and, therefore, we expect them to be more useful in the near term.  A prototypical heralding scheme uses a beamsplitter and a pair of single-photon detectors to generate heralded entanglement in a probabilistic fashion. To illustrate this, Fig. \ref{fig:theory} (a) shows a diagram of a simple two-node network.  Two dilution refrigerators with superconducting qubit chips are separated by kilometer-scale distances and connected by SMF-28 optical fiber.  In each fridge, an EO converter connects the microwave and optical domains by acting as a probabilistic source of entangled microwave-optical photon pairs.  When a pair is created, the microwave photon is sent to the superconducting qubit causing it to transition from an initial state $\ket{\psi_0}$ to a different state $\ket{\psi_1}$, and the optical photon is sent to the beamsplitter and single-photon detectors.  Because the beamsplitter erases the which-path information, the detection of an optical photon on either detector will project the two qubits into an entangled state $\ket{\psi} = \frac{1}{\sqrt{2}}\left(\ket{\psi_0}\ket{\psi_1} + \ket{\psi_1}\ket{\psi_0}\right)$. 

The present demonstration paves the way towards such future experiments of long-distance entanglement generation. We characterize the performance of the device primarily as a microwave-to-optical transducer. The integration of tunable filters and a single-photon detector into our experiment is an important step towards developing a source of entangled microwave-optical photon pairs. 

\subsection{Basic theory of operation}
Our electro-optic converter consists of a triply-resonant system utilizing two optical modes and one microwave mode, similar to the proposals in \cite{Soltani2017} and \cite{Gevorgyan2020}. Figs. \ref{fig:theory} (b) and (c) illustrates the device operation. Two nominally identical racetrack resonators evanescently couple to form optical modes.  After diagonalizing the Hamiltonian in the coupled-mode basis (see Supplement 1, section A), the interaction Hamiltonian for this triply-resonant system is given by 

\begin{equation}\label{eq:Hint_maintext}
\begin{aligned}
    \hat{H}_{\textrm{int}}/\hbar = g_o(\aop\bopd\cop + \aopd\bop\copd)
    \end{aligned}
\end{equation}
where $g_0$ is the EO coupling rate, and $\aop$, $\bop$, and $\cop$ are the annihilation operators for the symmetric optical mode, anti-symmetric optical mode, and microwave mode, respectively. Placing the optical pump on mode $a$ results in state transfer between optical mode $b$ and the microwave mode, while pumping mode $b$ allows for entangled photon pair generation in modes $a$ and $c$. 
When the two optical modes are separated by the microwave resonator frequency, the pump laser is tuned to the frequency of mode $a$ and the microwave drive is centered on mode $c$, the on-chip microwave-to-optical photon number conversion efficiency  is given by \cite{Tsang2011} (see Supplement 1, section B)
\begin{equation} \label{eq:eff_C_main_text}
    \eta =  \frac{\kappa_{b,e} \kappa_{c,e}}{\kappa_b \kappa_c} \frac{4C}{(1+C)^2}
\end{equation}
where $\kappa_{m,e}$ and $\kappa_{m,i}$ are the extrinsic and intrinsic loss rates of mode $m$, $\kappa_{b} = (\kappa_{b,i} + \kappa_{b,e})$ is the total loss rate for the single-side coupled optical mode $b$, and $\kappa_{c} = (\kappa_{c,i} + 2\kappa_{c,e})$ is the total loss rate for the double-side coupled microwave mode. The cooperativity $C = 4g_0^2 n_a / \kappa_b\kappa_c$ is resonantly enhanced by mode $a$ where $n_a$ is the intracavity number of photons. Our decision to use a triply-resonant rather than doubly-resonant design is because the presence of the second optical mode increases the intracavity pump photon number by a factor of approximately $4\omega_{\mu}^2  / \kappa_\textrm{opt}^2$ compared to using a detuned pump with a single optical mode (e.g. in \cite{Witmer2020}). For our device, this gives a reduction in required pump power of two orders of magnitude that is of paramount importance for cryogenic operation and sideband filtering for photon detection.

When the cooperativity $C \ll 1$, as is the case in this work, the on-chip efficiency can be expressed as (see Supplement 1, section B)

\begin{equation}\label{eq:Efficiency_LowC_with_detuning_main_text}
\begin{aligned}
       \eta \approx g_o^2 \frac{\kappa_{a,e}}{\Delta_a^2+(\frac{\kappa_a}{2})^2} \frac{\kappa_{b,e}}{\Delta_b^2+(\frac{\kappa_b}{2})^2} \frac{\kappa_{c,e}}{\Delta_c^2+(\frac{\kappa_c}{2})^2}     \frac{P_p}{\hbar\omega_a}.
    \end{aligned}
\end{equation}
where the detunings $\Delta_a = \omega_a - \omega_p$, $\Delta_b = \omega_b - \omega_p - \omega_\mu$, and $\Delta_c = \omega_c - \omega_\mu$, and $\omega_p$ and $\omega_\mu$ are the optical pump and microwave drive frequencies, respectively. 
Eqn. (\ref{eq:Efficiency_LowC_with_detuning_main_text}) shows the importance of having high quality factors, accurate detunings, and a large interaction rate $g_o$. Because efficiency is maximized when $\omega_b = \omega_a + \omega_c$, it is generally necessary to perform DC tuning of the optical modes.  Efficiency scales with the pump power ($P_p$) in the feed waveguide of the device, which places demands on the cryogenic optical fiber coupling into the chip which we solve by applying the packaging techniques shown in \cite{McKenna2019}. The cooling power of the dilution refrigerator also places limits on the pump power.  

Our device primarily uses the $r_{33}$ (contracted notation) component of the electro-optic tensor and transverse electric (TE) polarized fields along the extraordinary z-axis of the crystal with index $n_e$, so we can approximate $g_0$ as (see Supplement 1, section C) 
\begin{equation}\label{eq:g_o_r33_main_text}
\begin{aligned}
    \hbar g_o \approx \epsilon_o  n_{e}^4 r_{33}\int_{LN} dV \: e_a e_b^* e_c \cdot N_a N_b N_c.
    \end{aligned}
\end{equation}
We integrate the overlap of the fields inside the EO material. The fields $e_m(\textbf{r})$ are the three-dimensional electric field amplitudes, and the fields are normalized by their vacuum energies with $N_m = \sqrt{\hbar\omega_m / (2 \epsilon_o \int dV \sum_{ij}\epsilon_{ij}e_{mi} e_{mj}})$. Note that $N_c$ is proportional to the zero point voltage $V_{\textrm{zp}} = \sqrt{\hbar \omega_c/2C_\textrm{tot}}$, so the coupling is maximized by reducing the total circuit capacitance.   

If we want to use the device as an entangled pair source instead of a frequency converter, we put the optical pump on the higher frequency mode $b$.  In this case, the device probabilistically generates pairs of entangled microwave and optical photons with rate \cite{Tsang2011}
\begin{equation}
    R =  4C \frac{\kappa_{b,e} \kappa_{c,e}}{\kappa_b} ,
    \label{eqn:entangled_pairs}
\end{equation}
in the low cooperativity regime.

\begin{figure*} [htbp!]
\centering
\includegraphics[width=6in]{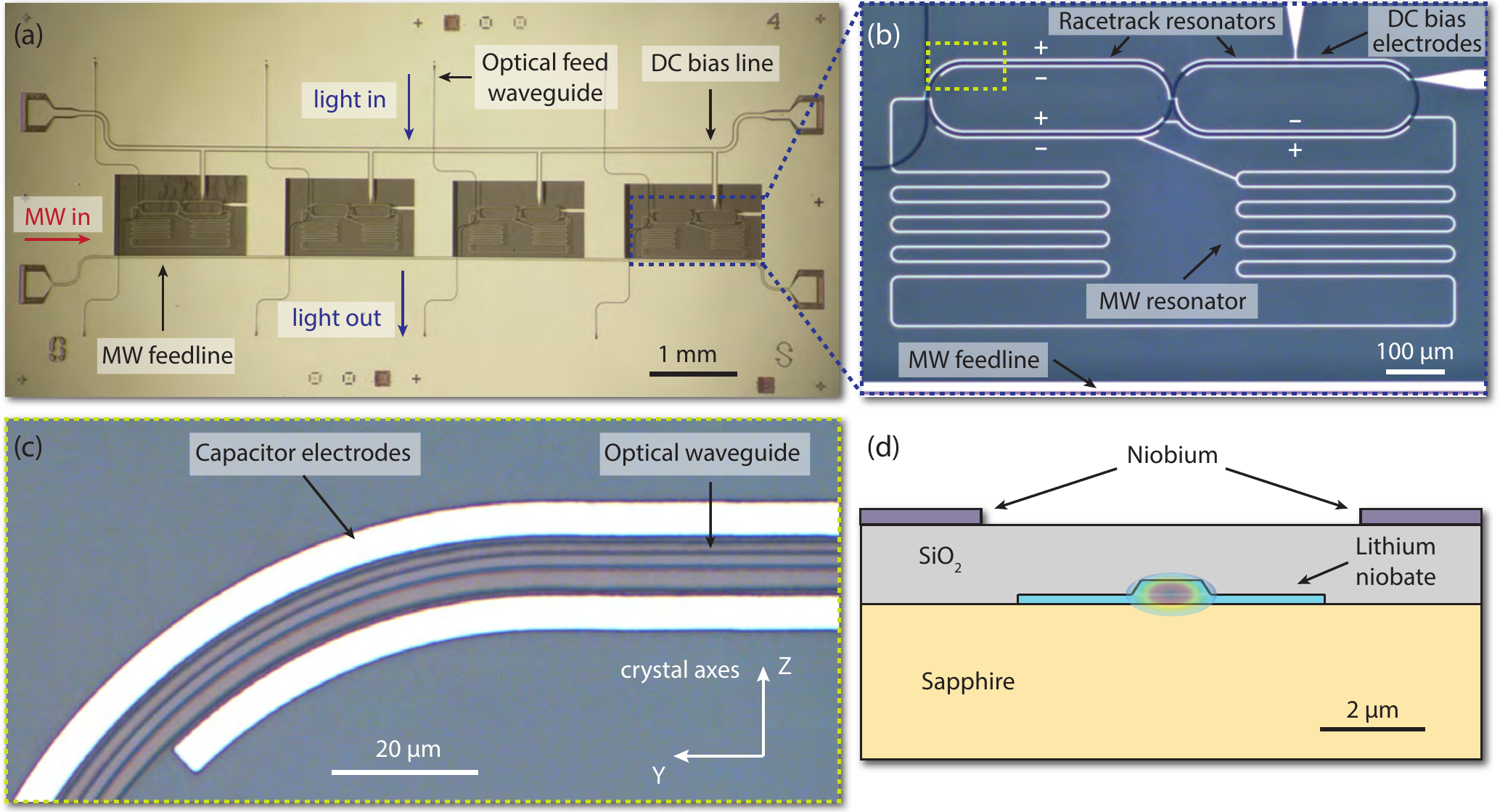}
\caption{(a) An image of the chip with four EO converter devices, before wirebonding and fiber gluing.  (b) An optical micrograph of the converter device, which consists of two coupled optical racetrack resonators and a quasi-lumped-element LC microwave resonator.  The $+$ and $-$ indicate the polarity of the modulation electrodes.  (c) A zoomed-in image showing the optical waveguide between the capacitor electrodes, with the LN crystal axes indicated.  (d) A diagram illustrating the cross-section geometry of the device (approximately to scale).  The exact waveguide dimensions can be found in Supplement 1.}
\label{fig:device}
\end{figure*}

\section{Lithium-niobate-on-sapphire (LiSa) platform}

\subsection{Motivation}
In order to achieve efficient on-chip microwave-to-optical conversion, it is necessary to use a material system that supports high optical and microwave quality (Q) factors as well as large electro-optic coupling.  Lithium niobate (LN) has long been a workhorse material for the telecommunications industry because of its large EO coefficient ($r_{33} \approx$ 31 pm/V \cite{Weis1985}) and low optical loss.  The recent commercial availability of thin-film lithium niobate wafers has enabled the development of nanofabricated LN resonators with optical Q factors well above 1 million \cite{Zhang2017,He2018,Desiatov2019,Zhang2019}.  These properties make LN an ideal candidate for microwave-to-optical conversion; however, the details of the material stack have a substantial impact on the device performance.  One commonly available thin-film LN platform, lithium-niobate-on-insulator (LNOI), consists of thin-film LN bonded to an LN or silicon substrate with a silicon dioxide buffer layer.  Although this platform has proven useful for classical modulator technology \cite{Wang2018}, the piezoelectric effect of the LN substrate can cause energy to be lost to acoustic radiation, presenting a significant loss channel for superconducting microwave resonators.  The silicon substrate variant also has challenges, since the interface between the oxide and silicon layers can exhibit large conductivity which also leads to microwave losses \cite{Wu1999}.  Lithium-niobate-on-silicon (LNOS), consisting of thin-film LN directly bonded to silicon, is an alternative stack which avoids these issues. However, since the refractive index of silicon is higher than LN, the LN film must be undercut in order to achieve optical guiding \cite{Jiang2019a,Jiang2020}, making the design and fabrication more challenging.

Lithium-niobate-on-sapphire, which we abbreviate as LiSa, consists of a thin LN film directly bonded to a sapphire substrate without any oxide layer, and it offers a promising alternative to both LNOI and LNOS for hybrid microwave/optical applications.   Sapphire is a commonly used microwave substrate with low dielectric and optical loss.

Unlike LN, sapphire is a centrosymmetric crystal and therefore not a source of piezoelectric loss.  The higher refractive index of LN relative to sapphire (approximately 2.2 compared to 1.7) makes it possible to confine light without the need for an intermediate buffer layer or suspended structures.

\subsection{Device overview}

The EO converter device is pictured in Fig. \ref{fig:device}.  The optical portion of the device consists of two coupled racetrack resonators, with the left resonator coupled to a feed waveguide. Grating couplers are used to couple light into and out of the chip from a pair of angle-polished fibers. The microwave resonator is a quasi-lumped-element LC circuit, with a meandering wire acting as the inductor and the modulator electrodes acting as the capacitor.  The microwave mode must have odd parity in the overlap region for the integral in Eqn. \ref{eq:g_o_r33_main_text} to be non-zero since modes $a$ and $b$ are the symmetric and anti-symmetric supermodes formed by coupling two resonators. To realize odd parity we arrange the electrodes as shown in Fig. \ref{fig:device} (b). The LC resonator is inductively coupled to a microwave coplanar waveguide (CPW) feedline.  Bias electrodes span a segment of the right optical resonator and are used to provide DC tuning to compensate for frequency mismatch.  The coupling between the optical resonators is carefully chosen to match the microwave resonator frequency when the optical modes are tuned to the symmetric operating point, i.e., when the modes are fully hybridized.

\subsection{Fabrication and packaging}

We fabricate the device using a LiSa wafer with 500 nm initial thickness of congruently grown X-cut lithium niobate on a C-cut sapphire substrate. The waveguides are defined with electron beam lithography (JEOL JBX-6300FS, 100-keV) in hydrogen silsesquioxane (HSQ) resist. The pattern is transferred to the lithium niobate with an argon ion mill etch that etches 300 nm of the lithium niobate. Next, we use photolithography and another argon ion mill etch to remove the remaining 200 nm of LN that would otherwise lie below the microwave circuit. We leave a 6 $\mu$m wide pedestal of LN around the waveguide.  A 1.5 $\mu$m thick oxide cladding is deposited on the die using plasma-enhanced chemical vapor deposition (PECVD) and annealed at 500 $^\circ C$.  Next, we RF sputter approximately 300 nm of niobium onto the die, pattern the microwave circuits using photolithography, and etch the niobium using a combination of argon ion milling and SF$_6$ reactive ion etching (RIE).  After dicing the die into smaller chips, the chips are then wirebonded into copper printed circuit boards (PCBs).  Angle-polished fibers are used to provide optical access to the device.  The fibers are aligned to the on-chip grating couplers and glued to the chips using the technique described in \cite{McKenna2019}.  The device is mounted at the 1 K still plate of a helium dilution refrigerator. Although the thermal occupancy of the microwave mode ($\approx 3$ photons) is not relevant to the current demonstration, future quantum experiments will require reducing the thermal occupancy by mounting the sample at a lower temperature location in the dilution refrigerator or thermalizing the microwave mode with radiative cooling \cite{Xu2020}.

\section{Electro-optic conversion results}

\begin{figure} [t!]
\centering
\includegraphics[width=\columnwidth]{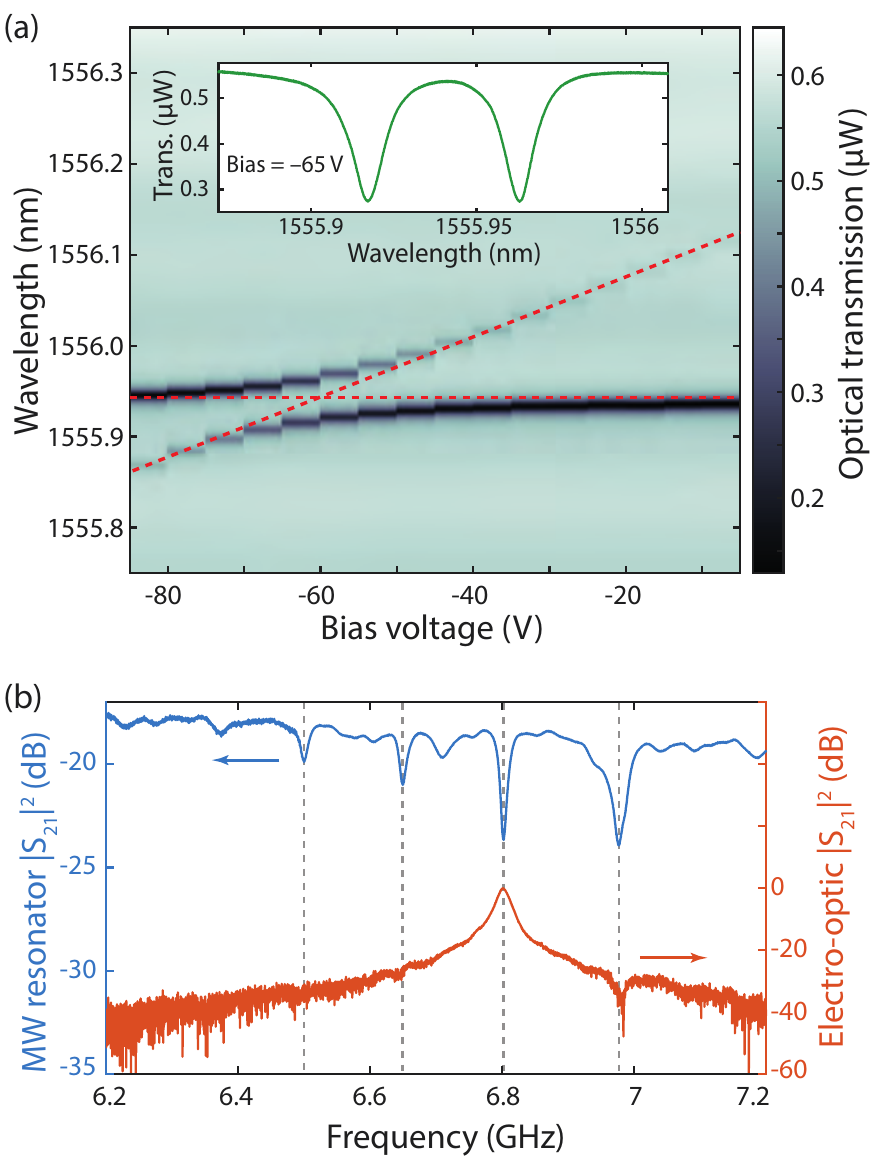}
\caption{(a) Optical spectra of a representative coupled racetrack device plotted vs. bias voltage.  The avoided crossing between the two modes is clearly visible.  Inset: An example spectrum showing the optical modes tuned to be fully hybridized.  In this case, the separation between the modes is approximately 46 pm.  (b) Microwave frequency characterization of the converter device.  The top curve (blue) shows an $S_{21}$ (transmission) measurement through the microwave feedline, taken using a vector network analyzer (VNA). The four resonances (marked with dashed lines) correspond to the four converter devices on the chip.  The bottom curve (red) is an electro-optic $S_{21}$ measurement showing the optical response on a high speed photoreceiver when the converter device is driven with a VNA.  The  $S_{21}$ is normalized to the peak EO response. For this measurement, light is going through the third device on the chip, which is why the EO response is strongly peaked at the frequency of the third microwave mode.  The data in this figure is taken from a different chip than the conversion data in Fig. \ref{fig:conversion_results}, but with nominally identical microwave circuits.}
\label{fig:linear_characterization}
\end{figure}

To prepare for EO conversion, the detuning between the optical modes must be adjusted to match the microwave frequency using the DC bias electrodes. The color plot in Fig. \ref{fig:linear_characterization} (a) shows optical transmission spectra as the bias voltage is swept, showing the avoided crossing between the two optical modes.  The inset shows an example spectrum where the modes are tuned to be fully hybridized. The optical modes used for the conversion process are undercoupled, and the splitting between the modes at the fully hybridized point is approximately 6.8 GHz. Device details are shown in table \ref{tab:parameters}.

The microwave $S_{21}$ (transmission) spectrum is shown in Fig. \ref{fig:linear_characterization} (b).  The four resonances correspond to the four LC resonators coupled to the feedline on the chip.  The microwave modes have total Q's in the range of 200--700.  The electro-optic response of the device can be measured by driving the microwave resonator and sending laser light through the feed waveguide to the coupled racetrack resonators.  The electric field on the capacitor modulates the light and creates an optical sideband which can be detected by sending the outgoing light to a high-speed photoreceiver. The EO response is highly peaked at 6.8 GHz since it is the third converter device on the chip which is being optically addressed.  From this data we conclude that the 3-dB conversion bandwidth is approximately 20 MHz.

\begin{figure*} [t!]
\centering
\includegraphics[width=7in]{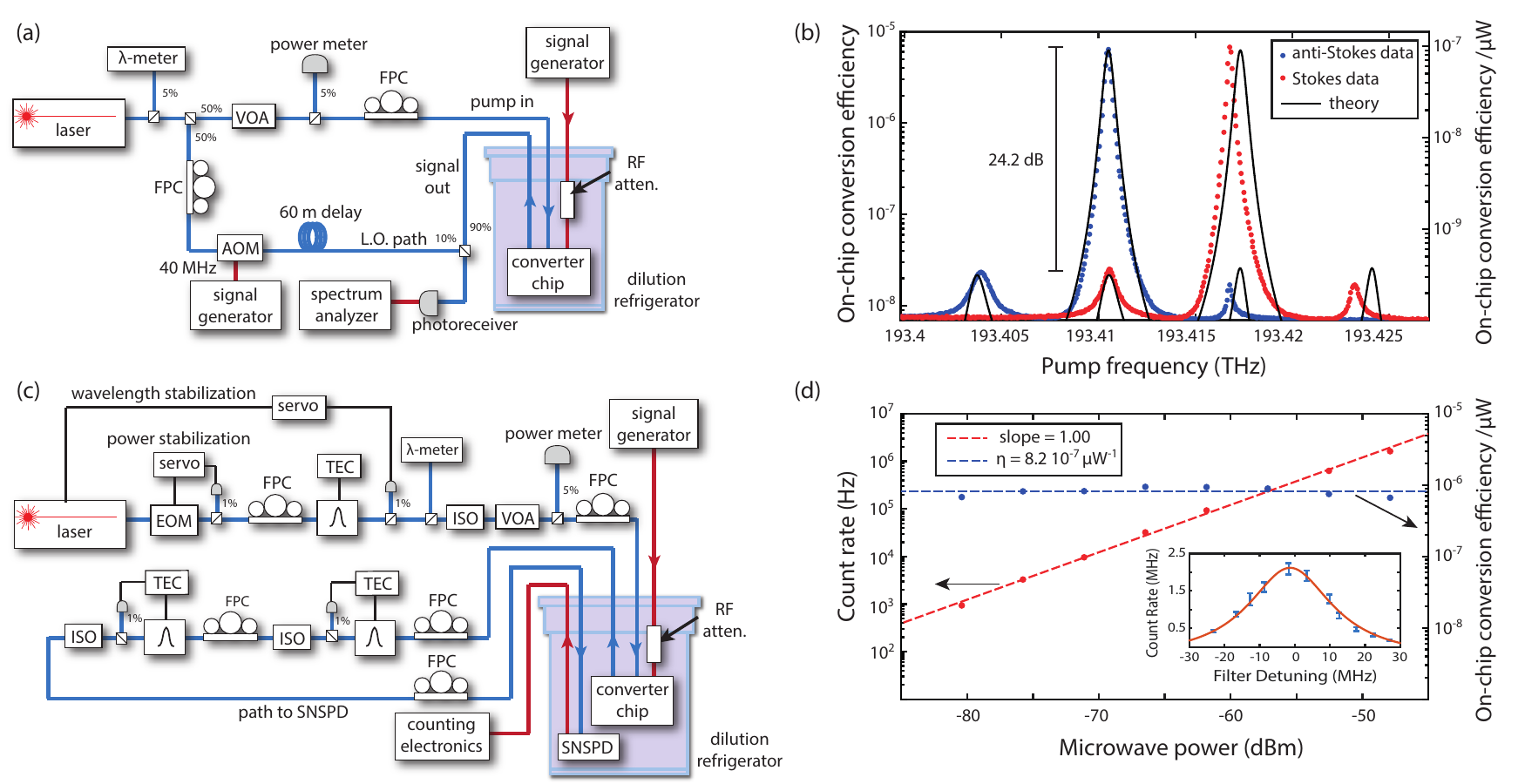}
\caption{(a) and (b) show the system diagram and measurement results for the heterodyne setup. (c) and (d) show the system diagram and measurement results for the SPD setup. (a) The system diagram for the heterodyne measurement setup. A 50/50 beamsplitter divides the light into two paths: one path pumps the device and the other path serves as the local oscillator (LO) path. The pump light drives the conversion process which results in Stokes and anti-Stokes sideband generation. The modulated light exits the device and combines with the LO with a 90/10 beamsplitter. An acousto-optic modulator (AOM) offsets the LO frequency by 40 MHz to allow for sideband discrimination in the beat signals of the high speed photoreceiver. (b) Experimental on-chip conversion efficiency versus the pump frequency. The device produces Stokes and anti-Stokes optical sidebands at $\omega_{p} - \omega_{\mu}$ and $\omega_{p} + \omega_{\mu}$, respectively, and the conversion efficiency into these sidebands is plotted versus pump frequency. The device resonantly enhances conversion while suppressing unwanted sideband generation. The black line shows the theoretical efficiency of the device. The theory curve uses measured parameters of the device and does not use any free fit parameters. The Stokes to anti-Stokes suppression ratio is 24.2 dB. (c) The system diagram for the single-photon detection setup that uses optical sideband filtering and a superconducting nanowire single-photon detector (SNSPD). The laser source is power stabilized using an electro-optic modulator (EOM) and is frequency stabilized to a high-finesse optical filter controlled by a thermo-electric cooler (TEC). The pump here is tuned to the anti-symmetric mode of the device (mode $b$). A variable optical attenuator (VOA) controls the amount of pump power to the device. The on-chip conversion process produces a Stokes sideband that is filtered by two high-finesse filters with a combined linewidth of around 30 MHz. The filters are temperature stabilized to the Stokes sideband frequency $\omega_{p} - \omega_{\mu}$. The SNSPD detects the filtered sideband signal and counting electronics measure the photon flux. (d) Experimental photon count rate versus the microwave drive frequency for a range of input microwave powers. The inset figure shows the count rate as a function of filter detuning. The count rate is maximum when the filter is tuned to $\omega_{p} - \omega_{\mu}$.  Other abbreviations: fiber polarization controller (FPC), optical isolator (ISO).}
\label{fig:conversion_results}
\end{figure*}

In order to demonstrate sideband selective conversion of microwave photons to optical photons, we use two different experimental techniques.  A heterodyne measurement shown in Fig. \ref{fig:conversion_results} (a) allows us to resolve the upper and lower sidebands with high frequency resolution and large dynamic range.  We send a resonant microwave tone and a pump laser to the device, and beat the modulated outgoing light with a frequency shifted local oscillator (LO) on a high-speed photoreceiver whose signal is fed to a realtime spectrum analyzer. By calibrating the measured spectra (see Supplement 1, section G for details) we can obtain the efficiency of conversion into the two sidebands as a function of the pump laser detuning (Fig. \ref{fig:conversion_results} (b)) and compare to the predicted curve from theory. The theory curve uses independently measured parameters such as the DC tuning response and Q factors and does not include any fit parameters from the heterodyne experiment.  Based on the measured efficiency we infer $g_0/2\pi=$ 1.2 kHz which very closely matches the value that we obtain from DC tuning measurements and is about $40\%$ lower than the value predicted from finite-element simulations (see Eqn. \ref{eq:g_o_r33_main_text}).

The maximum total on-chip photon number conversion efficiency is $6.6\times 10^{-6}$ dB ($3.9\times 10^{-7}$ off-chip), measured with 1.2 mW incident on the input grating coupler. The efficiencies reported generally have an uncertainty of $\pm 3.3$ dB which comes from an ambiguity in assigning the measured total losses to the input and output grating coupler losses. The insertion loss through both grating couplers is well known (24.4 dB), but decoupling the losses results in uncertainty. See supplement 1, section G for more details. The off-chip conversion efficiency includes the loss from the output grating coupler, but not the loss of the upstream microwave lines or downstream optical components.
The maximum on-chip conversion efficiency normalized to the pump power in the feed waveguide is $9.5\times 10^{-8} \: \mu W^{-1}$ ($5.7\times 10^{-9} \: \mu W^{-1}$ off-chip). Fig. \ref{fig:conversion_results} (b) shows the large advantage of incorporating an optical mode to resonate the pump. When the pump is set off-resonant with modes $a$ and $b$ by $\omega_c$, the converter operates with only a single optical resonance, and the efficiency is 24.2 dB lower than when utilizing both optical resonances. We define the selectivity as the ratio of the Stokes to anti-Stokes sideband efficiency with the pump tuned on resonance.
As viewed in Fig. \ref{fig:conversion_results} (b)., we achieve a selectivity of $24.2$ dB with the pump tuned on mode $a$, in good agreement with the theoretical prediction of 24.6 dB given our device parameters. Optical mode drift causes the peaks in efficiency to imperfectly match the theoretical predictions. The modes drifted by a larger amount towards the beginning of the experiment when the pump frequency was set to higher frequencies near mode $b$. We did not measure the photon conversion efficiency in the reverse direction (optical to microwave), but expect it to be the same based on the reciprocal nature of the device.

In our second measurement technique (illustrated in Fig. \ref{fig:conversion_results} (c)), the converted optical sideband photons are directly detected using a superconducting nanowire single-photon detector (SNSPD). A pair of cascaded, tunable Fabry-Perot filters (Micron Optics) are used to select the sideband of interest and provide approximately $110~\textrm{dB}$ suppression of pump light leaving the device.  The data is shown in Fig. \ref{fig:conversion_results} (d). We achieve a maximum on-chip conversion efficiency normalized to the pump power in the feed waveguide of $8.2\times 10^{-7} \: \mu W^{-1}$ ($4.9\times 10^{-8} \: \mu W^{-1}$ off-chip) with a pump power of 5.8 $\mu$W incident on the grating coupler. The background count rate is 4.8 kHz, due primarily to stray pump and environmental light reaching the SNSPD.   We note that there is a discrepancy between the efficiency per $\mu$W obtained using the two measurement techniques.  The two experiments were performed with very different pump powers incident on the input grating coupler (1.2 mW vs. 5.8 $\mu$W), and also during separate cooldowns of the fridge about one month apart.  We are still investigating the exact cause of the difference.

The incorporation of single-photon detectors in this experimental setup provides greater sensitivity and is an important step towards future quantum experiments such as heralded entanglement generation.
Based on our peak on-chip conversion efficiency, we estimate that this device operated without a microwave drive could generate entangled optical-microwave photon pairs at a rate of 20 Hz (see Eqn. \ref{eqn:entangled_pairs}).

A summary of the relevant device parameters is given in Table \ref{tab:parameters}.

\begin{table}
\centering
\begin{tabular}{|c|c|c|}
\hline
\textbf{Parameter}  & \textbf{Description}  & \textbf{Value} \\ \hline
$\omega_\textrm{\scriptsize opt}/2\pi$ &  Optical frequency & 193.411 THz    \\ \hline
$(\kappa_{a,i}, \kappa_{b,i})/2\pi$ & Intrinsic optical loss rate & (591, 466) MHz \\ \hline
$(\kappa_{a,e}, \kappa_{b,e})/2\pi$ &  Extrinsic optical loss rate & (206, 134) MHz \\ \hline
$(\kappa_a, \kappa_b)/2\pi$ &  Total optical loss rate &  (923, 600) MHz    \\ \hline 
 $2\mu/2\pi$ & Optical mode coupling & 6.8 GHz \\ \hline
$\omega_c/2\pi$ & MW resonance frequency &    6.801 GHz  \\ \hline
$\kappa_{c,i}/2\pi$ &  Intrinsic MW loss rate   &   12.8 MHz \\ \hline
$\kappa_{c,e}/2\pi$ & Extrinsic MW loss rate  &     4.4 MHz \\ \hline
$\kappa_c/2\pi$ & Total MW loss rate  &   21.6 MHz  \\ \hline
$g_o/2\pi$ & EO coupling rate  &   1.2 kHz  \\ \hline
\end{tabular}
\caption{Summary of measured device parameters.  Note that the optical resonances have single-sided coupling, so  $\kappa_\textrm{opt} = \kappa_{\textrm{opt}, i} + \kappa_{\textrm{opt},e}$, while the microwave resonator has two-sided coupling, so  $\kappa_c = \kappa_{c, i} + 2\kappa_{c,e}$. The microwave losses were measured with 43 dBm of microwave power to the device, so they represent high power Qs rather than single-photon Qs.
}
\label{tab:parameters}
\end{table}

\section{Other Design considerations}
There are a few design choices with the present device that are worth highlighting because of their broader applicability to microwave-to-optical converters.  

\begin{figure*} [htbp]
\centering
\includegraphics[width = 6in]{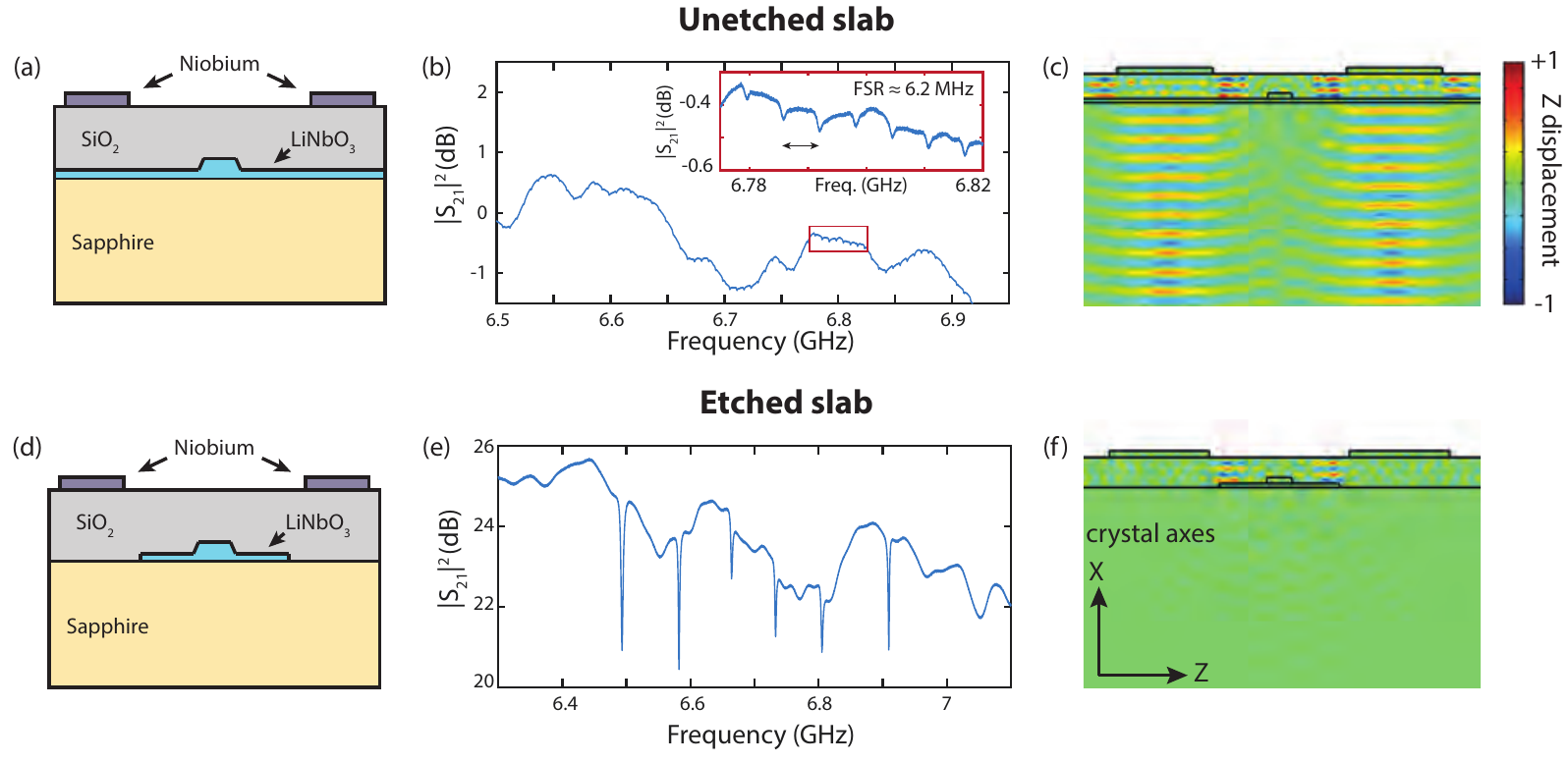}
\caption{The strategy used for reducing acoustic loss in the device.  (a) and (d) show a comparison of the device cross-section with and without the additional LN slab etch. (b) and (e) show microwave $S_{21}$ spectra from two nearly identical microwave LC resonators.  The device in (b) with an unetched slab has no clear resonances corresponding to the LC circuit, but there are a large number of small mechanical resonances corresponding to bulk acoustic waves (BAWs).  In contrast, the device in (e) with an etched slab has sharp resonances at the designed frequencies and no mechanical resonances are visible. (c) and (f) show COMSOL simulations of acoustic radiation from a device with an unetched slab and etched slab, respectively, simulated with an excitation frequency of 6.5 GHz. Etching the slab greatly reduces the excitation of shear waves.  The color shows the displacement in the crystal Z direction and the color scales in the two figures are the same.  }
\label{fig:acoustic_loss}
\end{figure*}

\subsection{Managing acoustic loss}

An issue we encountered while building the present converter device was the generation of acoustic waves by the microwave resonators.  Our initial devices had an LN slab which covered the entire chip underneath the microwave resonators (see Fig. \ref{fig:acoustic_loss}(a)).  These devices with an unetched slab had no clear microwave resonances at the designed frequencies; instead, we observed a large number of weakly coupled resonances with a free spectral range (FSR) of about 6.2 MHz.  We identify  these resonances as being due to acoustic shear waves in the sapphire substrate.  Transverse acoustic waves in sapphire have a velocity of approximately 6000 m/s \cite{Stanton2006}, and the sapphire substrate has a thickness of 500 $\mu$m, so the predicted FSR would be roughly 6 MHz, in line with what we observe.  The next iteration of microwave resonators was fabricated with a nearly identical design, except that the LN slab was etched everywhere except for a narrow 6 $\mu$m strip around the optical waveguides.  These devices were observed to have microwave resonances at the target frequencies with total Q factors of 2000--3000 (Fig. \ref{fig:acoustic_loss} (e)).  Note that one microwave resonator fabricated on the same chip had all of the LN completely etched away and had a Q factor in the same range, indicating that the Q is likely limited by another factor such as the dielectric losses in the oxide cladding rather than acoustic loss.  The dramatic difference in loss between the etched and unetched slab can be understood through simulations of the piezoelectric effect in our devices.  Fig. \ref{fig:acoustic_loss} (c) and (f) show a 2D COMSOL simulation of bulk acoustic waves generated when an AC voltage at 6.5 GHz is applied to the electrodes.  The dominant form of radiation is shear waves radiating downwards into the substrate.  These waves are generated by the vertical component of the electric field (along the LN X direction) directly underneath the electrodes, which interacts with the largest component of the LN piezoelectric tensor ($d_{15} \approx 7\times 10^{-11}$ C/N).  Since the electric field has this orientation only underneath the electrodes, removing the LN slab in this region is an effective way to reduce acoustic radiation.

\subsection{Choice of superconductor}
The choice of superconductor is very important to the converter operation.  Here we use niobium rather than aluminum because of its higher critical temperature ($\approx$ 8-9 K rather than $\approx$ 1 K) and much lower quasiparticle lifetime ($\approx$ nanoseconds compared to milliseconds).  This allows us to operate the device at the 1 K still plate of the dilution fridge and to use significantly higher pump powers.  Niobium titanium nitride (NbTiN) would likely be an even better choice than niobium because it has a similarly short quasiparticle lifetime and would also allow for the fabrication of high-kinetic inductance resonators to boost the EO coupling rate. Future work will explore the integration of high quality NbTiN films.  For more information on the effect of light on the microwave resonators refer to Supplement 1, section H.

\section{Outlook}

We have demonstrated sideband-resolved microwave-to-optical signal transduction in a dilution refrigerator environment.  Our device operates with an on-chip conversion efficiency of $6.6\times 10^{-6}$ and a large bandwidth of 20 MHz.

There are several straightforward ways in which the conversion efficiency of our transducer can be improved. First, we note that the microwave quality factor seems to be limited by the presence of the PECVD silicon dioxide cladding---test devices made with the same niobium film and similar circuit geometry, but without the oxide cladding, have intrinsic Q's above 100,000.  This is not surprising;  dielectric loss in amorphous materials often limits the Q factor of superconducting circuits, particularly at low microwave photon numbers where two-level system (TLS) effects become important \cite{Muller2019}.  Creating devices without the oxide should be straightforward in the LiSa platform by using airbridges~\cite{Fink2016} instead of oxide cladding, allowing the microwave circuits to be directly on the sapphire.  Additionally the optical Q factors of our resonators are similar with and without the cladding.  Other possibilities to improve the conversion efficiency include using higher impedance microwave resonators to enhance the EO coupling rate (for example by incorporating NbTiN nanowire kinetic inductors \cite{Hazard2019,Samkharadze2016,Shearrow2018}), increasing the optical Q by optimizing the electron beam lithography and ion mill etch, and improving fiber-to-chip coupling efficiency by either optimizing the grating coupler design or by switching to end-fire coupling using a lensed fiber \cite{Meenehan2014}.

It is also important to note that the optical modes in our experiment tended to drift, usually towards shorter wavelengths.  This drift depends on both the optical pump power and the applied bias voltage, and it is the limiting factor on how much pump power we can use in the experiment.  We believe this drift is due to a combination of photorefraction and optically induced conductivity in the LN, an effect which has been studied for several decades (see eg. \cite{Buse1997,Buse1997a,Hall1985,Schmidt1980}).  This effect is likely to be important for a broad class of thin film LN modulators---it is described in more detail in Supplement 1, section I and will be investigated carefully in future publications.  Mitigating this effect will greatly improve device performance by allowing us to send up to ten times more optical pump power to the device.

While there is still much work to be done in increasing conversion efficiency, this demonstration addresses some of the challenges in electro-optic photon conversion and highlights the potential of EO transducers to play a key role in future quantum networks.

\section*{Funding Information}
This work was supported by the US Government through the National Science Foundation under grant No. ECCS-1708734, Army Research Office  (ARO/LPS) CQTS program, through Airforce Office of Scientific Research (AFOSR) via (MURI No. FA9550-17-1-0002 led by CUNY), a DARPA Young Faculty Award (YFA). The authors wish to thank NTT Research Inc. for their financial and technical support. Part of this work was performed at the Stanford Nano Shared Facilities (SNSF) and Stanford Nanofabrication Facility (SNF). SNSF is supported by the National Science Foundation under award ECCS-1542152. A.S.N. acknowledges the support of a David and Lucile Packard Fellowship. R.V.L. acknowledges funding from the European Union's Horizon 2020 research and innovation program under Marie Sk\l{}odowska-Curie grant agreement No. 665501 with the research foundation Flanders (FWO). J.D.W. and P.A.A. acknowledge support from a Stanford Graduate Fellowship. E.A.W. acknowledges support by the Department of Defense (DoD) through the National Defense Science \& Engineering Graduate Fellowship (NDSEG) Program. R.N.P. acknowledges partial support from the NSF Graduate Research Fellowship Program under grant No. DGE-1656518.

\section*{Acknowledgments}
The authors would like to thank Dale Li, Hsaio-Mei (Sherry) Cho and Kent Irwin for assistance in depositing the niobium films, Pieter-Jan C. Stas for assistance with the counting circuit board layout, and Martin Fejer for helpful discussions about lithium niobate.  J.D.W. would like to thank Christopher Sarabalis for helpful discussions on the piezoelectric simulations.

\section*{Disclosures}

\noindent\textbf{Disclosures.} The authors declare no conflicts of interest.

\section*{Supplemental Documents}
See Supplement 1 for supporting content.

\newpage

\section{Supplementary Information}
\setcounter{equation}{0}
\renewcommand{\theequation}{S.\arabic{equation}}
\setcounter{figure}{0}  
\renewcommand\thefigure{S\arabic{figure}}    

\subsection{Derivation of diagonal coupled mode basis}

The Hamiltonian for our triply resonant system, in the basis of the uncoupled optical modes $\hat{a'}$ and $\hat{b'}$, is
\begin{equation}\label{eq:H_opt}
\begin{aligned}
    \hat{H}/\hbar= \omega_{a}'\aopdp\aopp + \omega_{b}'\bopdp\bopp + \omega_{c}\copd\cop + \mu(\aopp\bopdp + \aopdp\bopp) + \hat{H}_\textrm{int}/\hbar
\end{aligned}
\end{equation}
 where $\mu$ is the coupling rate between the left and right optical resonators and $\cop$ is the annihilation operator for the microwave mode.  The interaction part of the Hamiltonian is
 \begin{equation}
     \hat{H}_\textrm{int}/\hbar = 2g_V V_\textrm{zp}(\cop + \copd) \aopdp\aopp     \\ - g_V V_\textrm{zp}(\cop + \copd)
    \bopdp\bopp
 \end{equation}
where $V_\textrm{zp}$ is the zero-point voltage of the microwave resonator and  $g_V$ is the is the electro-optic tuning rate of the optical resonance frequency due to a single set of electrodes along one side of the racetrack resonator.  Note that the left racetrack mode $\aopp$ experiences twice the modulation of the right racetrack mode $\bopp$ because of our specific electrode configuration.  
 
 We can diagonalize the Hamiltonian in Eqn. \ref{eq:H_opt} by making a change of basis
\begin{equation}
\begin{aligned}
    \aop = \cos\theta~\aopp + \sin\theta~\bopp\\
    \bop = -\sin\theta~\aopp + \cos\theta~\bopp
\end{aligned}
\end{equation}
where the mixing angle $\theta$ is given by
\begin{equation}
    \tan2\theta = \frac{2\mu}{\Delta'} 
    \end{equation}
with $\Delta' = \omega_b' - \omega_a'$.  In the case when $\Delta' = 0$, $\aop$ and $\bop$ are the symmetric and anti-symmetric modes respectively. In this new basis the Hamiltonian becomes
\begin{equation}
\begin{aligned}
    \hat{H}/\hbar = \omega_{a}\aopd\aop + \omega_{b}\bopd\bop + \omega_{c}\copd\cop +  \hat{H}_\textrm{int}/\hbar
\end{aligned}
\end{equation}
where $\omega_a = \omega_a' - \textrm{sign}(\Delta')\sqrt{\mu^2 + \Delta'^2/4}$ and $\omega_b = \omega_b' + \textrm{sign}(\Delta')\sqrt{\mu^2 + \Delta'^2/4}$.  In the new basis, the interaction part of the Hamiltonian becomes
\begin{equation}
\begin{aligned}
    \hat{H}_\textrm{int}/\hbar = g_V V_\textrm{zp}(\cop + \copd) [ (2\cos^2\theta - \sin^2\theta)\aopd\aop \\+ (2\sin^2\theta - \cos^2\theta)\bopd\bop 
    \\+ 3\sin\theta\cos\theta(\aop\bopd + \aopd\bop)].
\end{aligned}
\end{equation}
In this work, we operate the device with $\Delta' = 0$, and only the last term in the square brackets contributes to coupling between the symmetric and anti-symmetric modes. We can simplify the interaction as 
\begin{equation}
\begin{aligned}
    \hat{H}_\textrm{int}/\hbar = \frac{3}{2} ~g_V V_\textrm{zp}(\aop\bopd\cop + \aopd\bop\copd)
\end{aligned}
\end{equation}
where we only keep resonant terms. The coupling rate $g_o$ is then $3g_V V_{\textrm{zp}}/2$.

\subsection{Derivation of conversion efficiency}
Adding in the optical and microwave drives, the system Hamiltonian becomes
\begin{equation}\label{eq:H}
\begin{aligned}
    \hat{H}/\hbar = \omega_{a}\aopd\aop + \omega_{b}\bopd\bop + \omega_{c}\copd\cop + g_o(\aop\bopd\cop + \aopd\bop\copd) \\ + \epsilon_p(\aop e^{i\omega_{p}t} + \aopd e^{-i\omega_{p}t}) + \epsilon_{\mu}(\cop e^{i\omega_{\mu}t} + \copd e^{-i\omega_{\mu}t}).
\end{aligned}
\end{equation}
Here $\aop$, $\bop$, and $\cop$ are the annihilation operators for the symmetric optical mode, antisymmetric optical mode, and microwave mode, respectively; $g_o$ is the undriven coupling rate between the modes, and $\epsilon_{p} = \sqrt{\kappa_{a,e}P_p/\hbar\omega_p}$ and $\epsilon_{\mu} = \sqrt{\kappa_{c,e}P_{\mu}/\hbar\omega_\mu}$ are the input field amplitudes of the red-detuned optical pump and microwave drive, respectively. After rotating out the time-dependence of the optical and microwave drives, the Heisenberg equations of motion read
\begin{equation}\label{eq:EOMs}
\begin{aligned}
    &\frac{d}{dt}\aop = (-i\Delta_a - \frac{\kappa_a}{2})\aop -ig_o\bop\copd -i\epsilon_p \\ 
    &\frac{d}{dt}\bop = (-i\Delta_b - \frac{\kappa_b}{2})\bop -ig_o\aop\cop \\
    &\frac{d}{dt}\cop = (-i\Delta_c - \frac{\kappa_c}{2})\cop -ig_o\aopd\bop -i\epsilon_{\mu}, 
\end{aligned}
\end{equation}
where $\Delta_a = \omega_a-\omega_p$, $\Delta_b = \omega_b-\omega_p-\omega_{\mu}$, and $\Delta_c = \omega_c - \omega_{\mu}$. 

We solve these equations in steady state by setting the time derivatives in Eqn. \ref{eq:EOMs} to zero and solving for the photon conversion efficiency $\eta = |b_{\textrm{out}}/c_{\textrm{in}}|^2$ where $|b_{\textrm{out}}|^2=\kappa_{b,e}|b|^2$ is the output flux of generated optical photons and $|c_{\textrm{in}}|^2=|\epsilon_{\mu}|^2/\kappa_{c,e}$ is the input flux of microwave photons. We also assume we only pump on mode $a$ and there is no backaction conversion from modes $b$ and $c$ to mode $a$, which is equivalent to setting the term $-ig_o\bop\copd$ in Eqn. \ref{eq:EOMs} to zero.

To find the on-chip conversion efficiency of coherent fields, we displace our operators $\aop \xrightarrow{} \aop+\bar{a}$ and then take the mean fields which is equivalent to converting operators to complex field amplitudes as $\aop \xrightarrow{} \bar{a}$, $\bop \xrightarrow{} \bar{b}$, and $\cop \xrightarrow{} \bar{c}$. The on-chip efficiency is

\begin{equation}\label{eq:FullEfficiency}
\begin{aligned}
    \eta = \frac{G^2}{|1 + \frac{G^2}{D_{b}D_{c}}|^2}  \frac{\kappa_{b,e}\kappa_{c,e}}{|D_{b}|^2 |D_{c}|^2 }
    \end{aligned}
\end{equation}
where for convenience we have packaged the denominator terms as $D_m = -i\Delta_{m} - \frac{\kappa_m}{2}$ and the square of the driving-enhanced coupling rate $G$ is
\begin{equation}\label{eq:BigGSquared}
\begin{aligned}
    G^2=g_o^2 \bar{a}^2 = g_o^2 \frac{\kappa_{a,e}}{|D_a|^2}\frac{P_p}{\hbar\omega_a}
    \end{aligned}
\end{equation}

When the signals and modes are perfectly tuned in frequency, $\Delta_a=\Delta_b=\Delta_c=0$, and the efficiency becomes the familiar:
\begin{equation}\label{eq:FullEfficiency_zero_detuning}
\begin{aligned}
    \eta = \frac{4C}{(1+C)^2} \frac{\kappa_{b,e}\kappa_{c,e}}{\kappa_{b}\kappa_{c}}
    \end{aligned}
\end{equation}
where the cooperativity $C$ is $4G^2/\kappa_b\kappa_c$. Our device operates in the low cooperativity regime, where $(1+C)^2 \approx 1$ and the conversion efficiency with zero detuning is $\eta \approx 4C \: \kappa_{b,e}\kappa_{c,e}/\kappa_b\kappa_c$. Now re-introducing the impact of detuning into the low cooperativity efficiency expression yields
\begin{equation}\label{eq:Efficiency_LowC_with_detuning}
\begin{aligned}
       \eta \approx g_o^2 \frac{\kappa_{a,e}}{\Delta_a^2+(\frac{\kappa_a}{2})^2} \frac{\kappa_{b,e}}{\Delta_b^2+(\frac{\kappa_b}{2})^2} \frac{\kappa_{c,e}}{\Delta_c^2+(\frac{\kappa_c}{2})^2}     \frac{P_p}{\hbar\omega_a}.
    \end{aligned}
\end{equation}

Eqn. \ref{eq:Efficiency_LowC_with_detuning} shows that non-zero detuning limits efficiency, so it is important that the hybridized modes $a$ and $b$ are separated by $\omega_{\mu}$. The efficiency also grows linearly with pump power, but heat dissipation in the dilution fridge environment and charge induced device drift place practical limits on pump power in this experiment. Assuming perfect detuning $\Delta_m=0$ and critical coupling $\kappa_m = 2 \kappa_{m,i}$, the low-cooperativity efficiency is
\begin{equation}\label{eq:Efficiency_LowC_no_detuning_CC}
\begin{aligned}
       \eta \approx \frac{8 g_o^2}{\kappa_{a,i}\kappa_{b,i}\kappa_{c,i}}
       \frac{P_p}{\hbar\omega_a}.
    \end{aligned}
\end{equation}
Eqn. \ref{eq:Efficiency_LowC_no_detuning_CC} clearly shows that to maximize efficiency, we wish to minimize intrinsic losses of all resonances, maximize $g_o$, and maximize the pump power to drive the conversion process.

\subsection{Derivation of coupling rate $g_o$}
\label{sec:g_o_derivation}
The three wave mixing interaction between optical mode $a$, optical mode $b$, and microwave mode $c$ is described by the Hamiltonian
\begin{equation}\label{eq:Hint}
\begin{aligned}
    \hat{H}_{\textrm{int}}/\hbar = g_o(\aop\bopd\cop + \aopd\bop\copd).
    \end{aligned}
\end{equation}

The first term in Eqn. \ref{eq:Hint} describes the conversion of a pump photon at mode $a$ with an input microwave photon at mode $c$ to produce an optical photon at mode $b$ with the conservation of energy requirement that $\omega_b = \omega_a + \omega_c$. The second term describes the decay of a photon in mode $b$ into a photon in mode $a$ and microwave photon in mode $c$ with the same conservation of energy requirement. The interaction Hamiltonian allows for quantum state transfer between microwave and optical photons (optical pump centered on mode $a$) or quantum entanglement between microwave and optical photons (optical pump centered on mode $b$) with enhanced efficiency due to ability to resonate the pump. 

To find an expression for $g_o$, we must find the interaction energy due to the microwave electric field modulating the index of refraction of the optical resonator due to the electro-optic effect. The electro-optic effect is usually expressed as a change in $1/\epsilon$ for an applied microwave field, so the electro-optic interaction energy is most conveniently expressed in terms of the inverse permittivity matrix $\boldsymbol{\eta}$ and the optical displacement field $\hat{D}_o$ as
\begin{equation}\label{eq:Hint_DD}
\begin{aligned}
    \hat{H}_{int}/\hbar & = \frac{1}{2\epsilon_o}\int_{LN} dV \sum_{ij} \Delta \eta_{ij} \hat{D}_{oi} \hat{D}_{oj} \\ & = \frac{\epsilon_o}{2} \int_{LN} dV\sum_{ij} \Delta \eta_{ij} \epsilon_{ii} \epsilon_{jj} \hat{E}_{oi} \hat{E}_{oj}.
    \end{aligned}
\end{equation}
We assume the fields are aligned to the crystal coordinates, and the volume integral is over the lithium niobate region. Now we substitute in the EO $r$ tensor as $\Delta\eta_{ij}=\sum_{ijk}r_{ijk} \hat{E}_{\mu k}$ and the microwave field as $\hat{E}_{\mu} = (e_c\cop + h.c)$ and the optical field as $\hat{E}_o = (e_a \aop + e_b \bop + h.c.)$ to form the expression
\begin{equation}\label{eq:Hint_uuu}
\begin{aligned}
    \hat{H}_{\textrm{int}}/\hbar = \epsilon_o \int_{LN} dV \: \sum_{ij} \epsilon_{ii}\epsilon_{jj} r_{ijk}(e_{ai} e_{bj}^* e_{ck} \: \aop\bopd\cop + h.c.) \cdot N_a N_b N_c
    \end{aligned}
\end{equation}
where only terms which conserve energy are kept. The complex vector electric fields $e_m(\textbf{r})$ are normalized to their zero point energies with the coefficients $N_m = \sqrt{\hbar\omega_m / (2 \epsilon_o \int dV \sum_{ij}\epsilon_{ij}e_{mi} e_{mj}})$. 
From Eqn. \ref{eq:Hint_uuu}, we identify $\hbar g_o$ as
\begin{equation}\label{eq:g_o}
\begin{aligned}
    \hbar g_o = \epsilon_o \int_{LN} dV \: \sum_{ijk} \epsilon_{ii}\epsilon_{jj} r_{ijk} e_{ai} e_{bj}^* e_{ck} \cdot N_a N_b N_c.
    \end{aligned}
\end{equation}
Our device primarily uses the $r_{33}$ (contracted notation) component of the electro-optic tensor and TE polarized fields along the extraordinary z-axis of the crystal with index $n_e$, so we can simplify Eqn. \ref{eq:g_o} as 
\begin{equation}\label{eq:g_o_r33}
\begin{aligned}
    \hbar g_o \approx \epsilon_o  n_{e}^4 r_{33}\int_{LN} dV \: e_a e_b^* e_c \cdot N_a N_b N_c.
    \end{aligned}
\end{equation}

\subsection{Equivalent expression for $g_o$}
The derivation of the coupling rate $g_o$ in section \ref{sec:g_o_derivation} relies on the change in energy of the optical field in the material as a result of the microwave field perturbing the permittivity. This ultimately results in $g_o$ expressed with the EO $r$ tensor in Eqn. \ref{eq:g_o}. An equivalent formulation expresses the interaction as the energy caused by the second-order nonlinear polarization induced in the material by the electric fields: 
\begin{equation}\label{eq:Hint_polarization}
\begin{aligned}
    \hat{H}_{\textrm{int}}/\hbar = \frac{1}{3} \int_{LN} dV \: \hat{P}^{\textrm{(2)}} \cdot \hat{E} = \frac{\epsilon_o}{3} \int_{LN} dV \: \sum_{ijk} \chi_{ijk}^{\textrm{(2)}}\hat{E}_i \hat{E}_j \hat{E}_k.
    \end{aligned}
\end{equation}
The electric field includes microwave and optical fields as $\hat{E} = e_a\aop + e_b\bop + e_c\cop + h.c.$ Plugging in and normalizing the fields to their vacuum energies results in the expression
\begin{equation}\label{eq:g_o_Chi2}
\begin{aligned}
    \hbar g_o = 2 \epsilon_o \int_{LN} dV \sum_{ijk} \chi_{ijk}^{\textrm{(2)}} e_{ai} e_{bj}^* e_{ck} \cdot N_a N_b N_c.
    \end{aligned}
\end{equation}
Finally, comparing Eqn. \ref{eq:g_o_Chi2} with Eqn. \ref{eq:g_o} shows that $\chi_{ijk}^{\textrm{(2)}} = \epsilon_{ii}\epsilon_{jj}r_{ijk}/2$.
\subsection{Added thermal noise}
Although the present experiment is not sensitive to the effects of thermal noise in our microwave circuit, considering the impact of thermal noise is important for future quantum experiments.  The thermal occupation $n_c$ of the microwave mode is given by the Bose-Einstein distribution
\begin{equation}
    n_c = \frac{1}{\exp(\hbar \omega_\mu / k_B T) - 1}.
\end{equation}
Because thermal microwave photons will be transduced to optical photons in the same way as signal photons coming from a qubit, we need to reduce the thermal occupation to be much less than one in order to prevent thermal noise from impacting the transduction fidelity.  If we assume that the microwave mode of our device is well thermalized to the still plate temperature of $\approx$ 1 K, then we expect a thermal occupation of $n_c \approx 3$.  Mounting the device in a colder location in the dilution refrigerator could help reduce $n_c$; for example, at the 100 mK plate $n_c$ would be reduced to about 0.04, and thermalizing at the mixing chamber ($T = 10$ mK) would eliminate thermal noise completely ($n_c \approx 8\times 10^{-15}$).  However, reducing $n_c$ in this way is difficult in practice due to limited cooling power at these colder stages ($< 10 \mu W$) which is incompatible with the current amount of pump light needed to drive the conversion process. Ref. \cite{Xu2020} shows a radiative cooling scheme that could be used to reduce $n_c$ while keeping the sample mounted at the 1 K stage.

\subsection{Optical and microwave co-design}

\begin{figure} [t!] 
\centering
\includegraphics[width=\columnwidth]{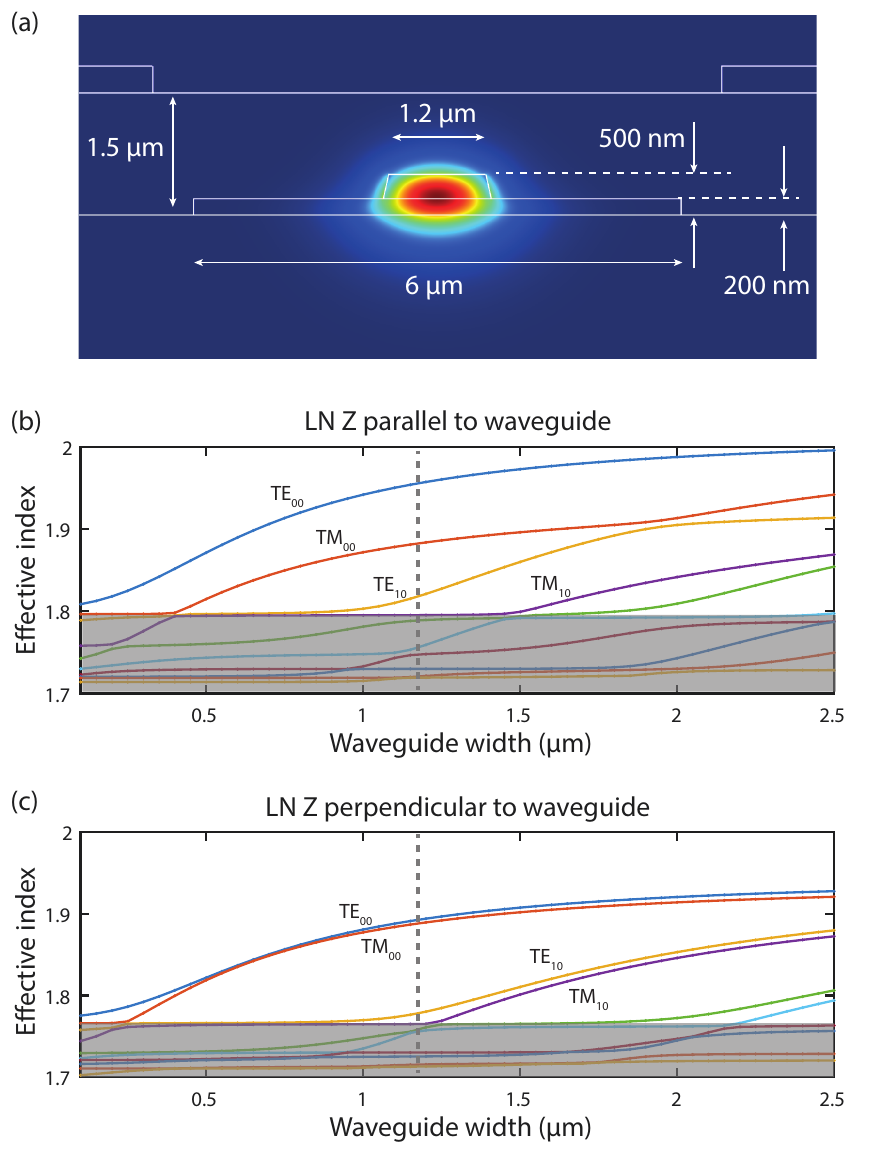}
\caption{(a) COMSOL simulation showing the fundamental quasi-TE optical mode of the LiSa waveguide (at 1555 nm) and the waveguide dimensions.  The color shows the electric field norm. (b) and (c) show the optical mode effective indices vs. waveguide width when the LN crystal Z axis is parallel and perpendicular to the light propagation direction, respectively.  When Z is parallel to the waveguide, the TE$_{00}$-like and TM$_{00}$-like modes are clearly separated, while for the perpendicular direction these modes are nearly degenerate.  The grey dashed lines show the waveguide width of 1.2 $\mu$m used in this device, and the grey solid regions show unguided modes which lie below the light line for the LN slab. }
\label{fig:optical_design}
\end{figure}

The LiSa waveguide geometry was designed using finite-element simulations (COMSOL); the waveguide dimensions and fundamental quasi-TE mode profile are shown in Fig. \ref{fig:optical_design} (a).  Figs. \ref{fig:optical_design} (b) and (c) show the optical mode effective indices vs. waveguide width when the LN crystal Z axis is parallel and perpendicular to the light propagation direction, respectively.  We choose a waveguide width of 1.2 $\mu$m to achieve maximum confinement of the optical mode in the LN while reducing the impact of higher order spatial modes.  Due to the strong birefringence of LN ($n_o = 2.21$, $n_e = 2.14$), the optical mode indices depend strongly on waveguide orientation.  When Z is parallel to the waveguide, the TE$_{00}$-like and TM$_{00}$-like modes are clearly separated, while for the perpendicular direction these modes are nearly degenerate.  

The optical and microwave resonators in our converter device are designed together in order to maximize the electro-optic cooperativity.  Optical racetrack resonators are used instead of simple ring resonators in order to maximize the fraction of the waveguide which is perpendicular to the LN crystal Z axis.  The bending radius of these racetracks is chosen to be large enough to avoid optical bending loss but not much larger, since increasing the racetrack size also increases the capacitance of the microwave resonator and therefore lowers the impedance.  We chose a bending radius of 72 $\mu$m and a straight section length of 350 $\mu$m, which approximately optimizes the coupling rate $g_0$.  The resulting free spectral range (FSR) for the racetrack resonators is approximately 900 pm.

\begin{figure} [htbp]
\centering
\includegraphics[width=\columnwidth]{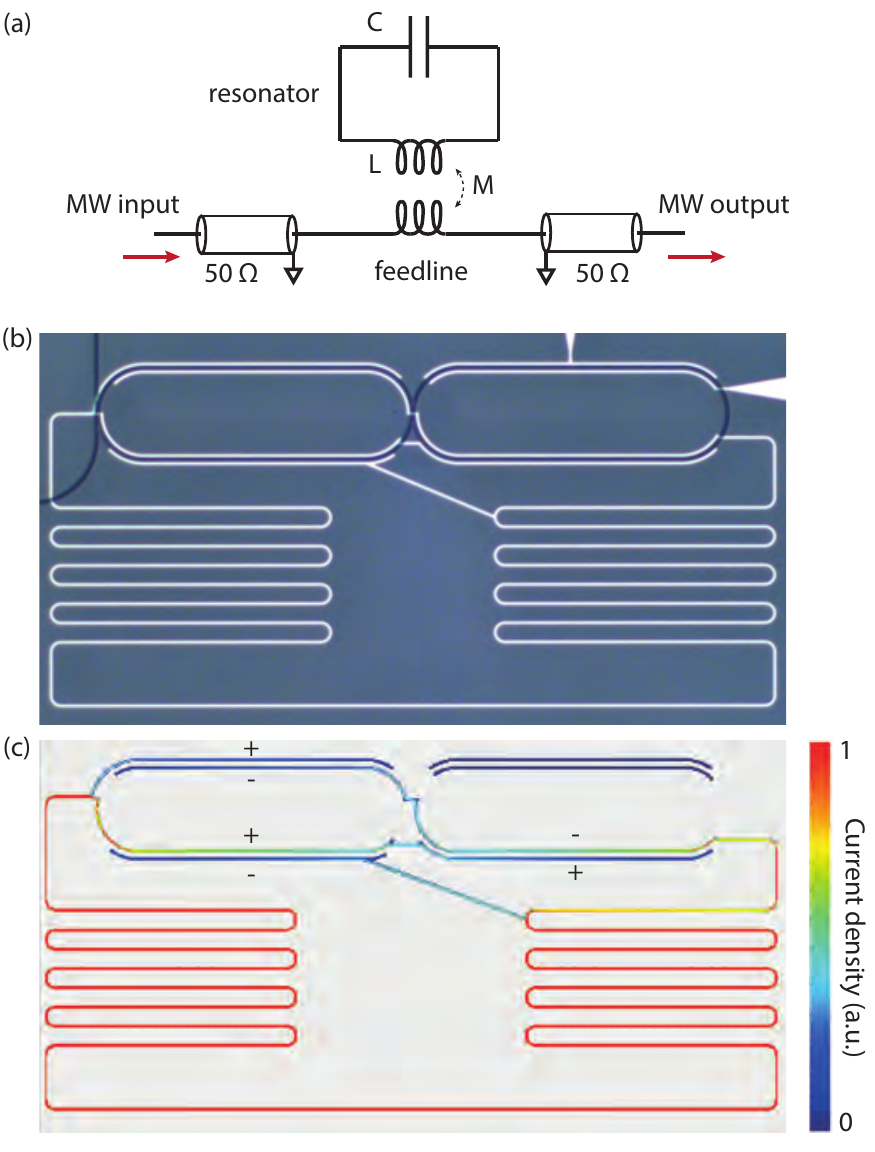}
\caption{(a) Lumped-element equivalent circuit of the microwave resonator.  (b) Optical micrograph of the microwave resonator.  (c) Sonnet simulation showing the current density of the resonator and the polarity of the modulation electrodes.}
\label{fig:MW_resonator}
\end{figure}

The microwave resonator is designed to be a quasi-lumped-element LC circuit, where the meandering wire provides the inductance and the modulating electrodes provide the capacitance (see Fig. \ref{fig:device} (a)).  The coupled racetracks are modulated by the microwave resonator on three of the four straight sections, while the fourth is reserved for DC biasing (see Fig. \ref{fig:device} (b)).  Note that the electrode geometry was designed so that the direction of the electric field between the electrodes switches direction between the left and right racetracks; this is necessary in order for the modulation to contribute to the EO coupling, rather than providing a common frequency shift.  A Sonnet simulation of the current distribution in the microwave resonator is shown in Fig. \ref{fig:MW_resonator} (c).  The simulated impedance of the resonator is approximately 300 $\Omega$.

\subsection{Calibration of conversion efficiency}
To calibrate the photon number conversion efficiency, we need to accurately determine the microwave power input to the device and the optical sideband power leaving the device.  Based on measurements of cable loss, we estimate that the microwave signal experiences a total of 13 dB of attenuation between the RF signal generator and the device (2 dB from cables outside the fridge and 11 dB from the fridge input lines).  On the optical side, we calibrate the optical sideband power measurement by sending a known amount of sideband and frequency shifted LO power to the high-speed photoreceiver.  This produces a peak at 40 MHz on the real-time spectrum analyzer (RSA), and we integrate the power spectral density of the peak to determine the amount of microwave power incident on the RSA. The relationship between the optical sideband power $P_\textrm{sideband}$ incident on the detector and the microwave power $P_\textrm{MW}$ incident on the RSA is approximately linear, given by
\begin{equation}
    P_\textrm{MW} = G \cdot P_\textrm{sideband} \cdot P_\textrm{LO}
\end{equation}
where $P_\textrm{LO}$ is the LO power and $G$ is the gain factor determined in the calibration.  We found $G$ to be 1.02$\times10^4$ W$^{-1}$, measured with an LO power of 390 $\mu$W and a sideband power of 3.4 $\mu$W.  The insertion loss of the optical fibers in the fridge and the optical components in the setup are measured independently.  The total combined loss of the input and output grating couplers is determined via optical transmission measurements to be 24 dB.  However, it is difficult to accurately determine the individual losses of each grating coupler independently. Based on our previous experience with loss in these grating couplers, we expect the loss for each individual coupler to be between 9 and 15 dB.  While this does not effect the measured total \emph{off-chip} conversion efficiency or the \emph{on-chip per $\mu$W} conversion efficiency, it is the largest source of uncertainty in the \emph{on-chip} conversion efficiency and the \emph{off-chip per $\mu$W} efficiency.

\subsection{Microwave resonator response to optical pump}
\label{sec:Supp_heating}

\begin{figure} [t!] 
\centering
\includegraphics[width=\columnwidth]{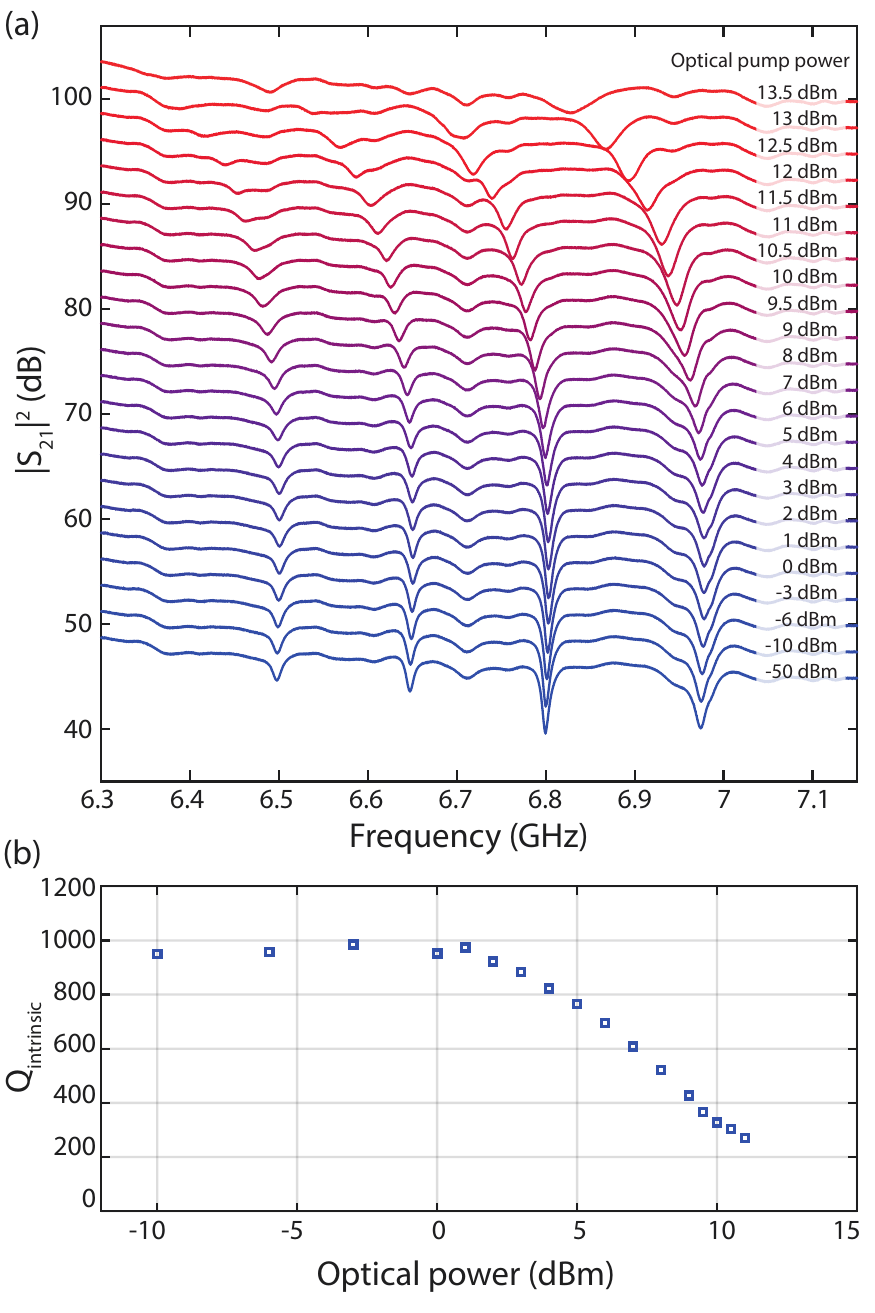}
\caption{(a) The effect of increasing optical pump power on the microwave transmission spectrum.  As expected from Mattis-Bardeen theory, the four modes experience a decrease in both frequency and Q factor.  (b) The intrinsic Q factor for the third microwave mode (bare frequency $\approx$ 6.8 GHz) vs. optical pump power into the fridge.  The microwave Q begins to degrade significantly for powers above about 3 dBm.  This data in this figure is taken from a different chip than the conversion data in Fig. 4, but with nominally identical microwave circuits.}
\label{fig:CW_heating}
\end{figure}

One of the major limiting factors for the microwave-to-optical conversion efficiency in this demonstration is the reduction of the microwave Q factor by the strong optical pump light.  Fig. \ref{fig:CW_heating} shows the change in the microwave spectrum with different amounts of CW pump light applied.  As the light heats up the chip, the microwave modes decrease in frequency and increase in linewidth, as predicted by Mattis-Bardeen theory \cite{Mattis1958,Zmuidzinas2012,Visser}.  The spectra did not significantly depend on whether or not the laser was resonant with one of the optical modes. We note that the niobium resonators here begin to exhibit increased loss at much higher optical pump powers compared to the aluminum resonators in our previous work \cite{Witmer2020}, which began to degrade with even 1 $\mu$W of optical power. This is likely due to the shorter quasiparticle lifetime and higher critical temperature of niobium compared to aluminum.  We can use the built-in resistive thermometer to measure the effect of light on the temperature of the 1 K still plate of the dilution refrigerator.  For the highest optical powers (13.5 dBm), we see the temperature increase from approximately 1.01 K to 1.25 K. The temperature change is moderated by the large cooling power at the still plate ($\approx$ 30 mW).  Since this relatively small temperature change of 240 mK cannot explain the large decrease in Q factor, we believe that the heating is largely localized on the chip.

\begin{figure*} [t!] 
\centering
\includegraphics[width=5in]{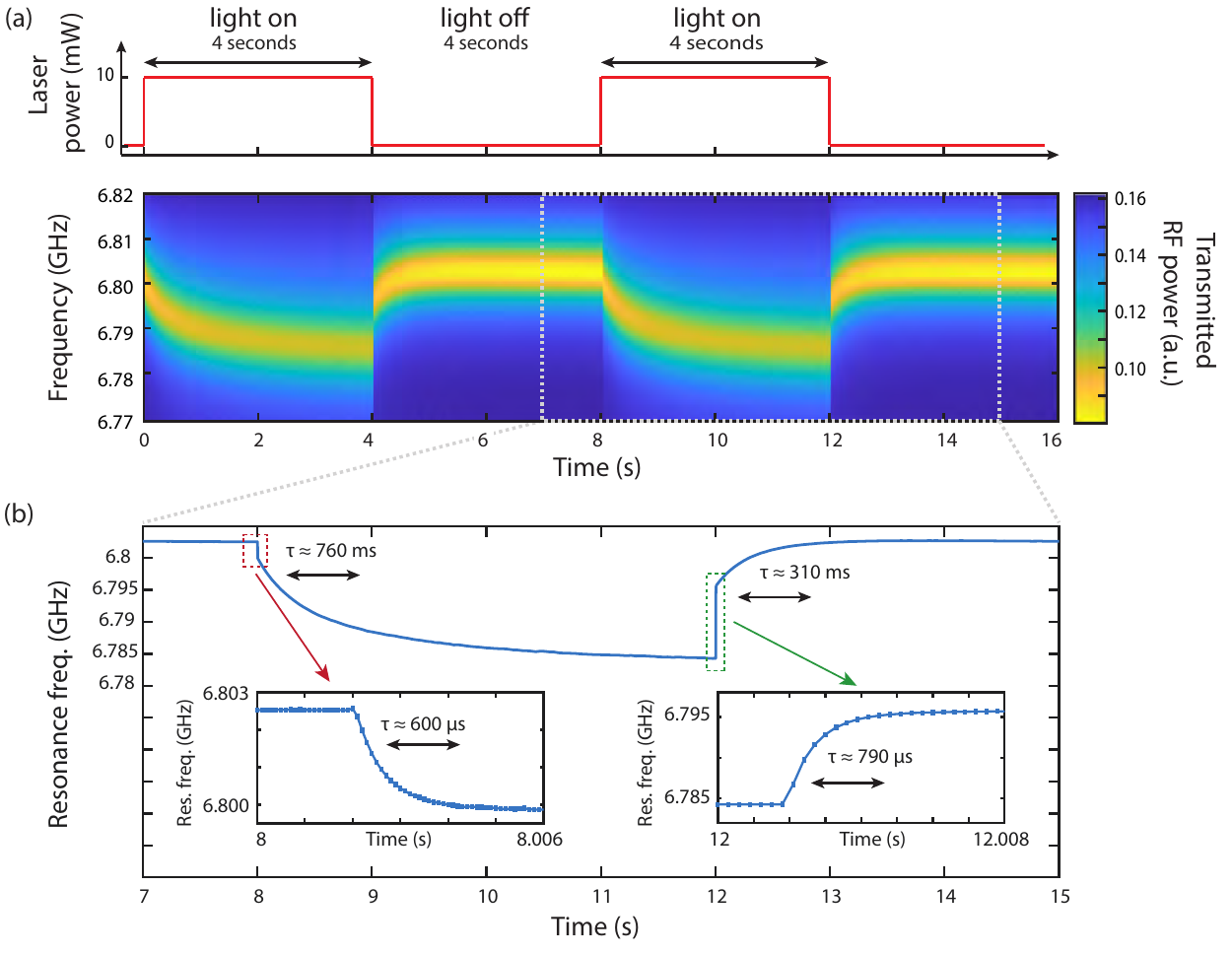}
\caption{Temporal dynamics of the microwave resonator response to optical pump light.  (a) A color plot of the microwave transmission as 10 mW of optical pump light is switched on and off for 4 seconds each.  As expected, the microwave mode broadens and decreases in frequency when the light is turned on, and returns to it's original position when the light is switched off.  (b) A plot of the microwave resonance frequency over time.  The frequency is extracted by fitting time slices from (a) to a Lorentzian.  Insets: zoomed-in plots showing the initial response to the light turning on/off in more detail. We observe two very distinct timescales in the microwave response.  The time constants $\tau$ given in the plot are from fitting to an exponential decay. This data in this figure is taken from a different chip than the conversion data in Fig. 4, but with nominally identical microwave circuits.}
\label{fig:pulsed_heating}
\end{figure*}

We also investigated the temporal dynamics of the microwave spectra, using an acousto-optic modulator (AOM) as a fast switch for the light.  We excite the microwave resonators with a continuous-wave (CW) tone at different frequencies and use a spectrum analyzer in IQ mode to measure the time-dependent amplitude and phase of the transmitted signal.  Fig. \ref{fig:pulsed_heating} shows the response of the device to 10 mW of light for 4 second intervals.  We observe two distinct time scales in the response, one in the range of hundreds of microseconds, and the other hundreds of milliseconds.  Both of these timescales are much too long to correspond to the direct optical generation of quasiparticles (the maximum quasiparticle lifetime in niobium is approximately 1 ns \cite{Visser,Lobo2005}).  One possibility is that the shorter timescale corresponds to local heating of the sapphire chip, while the longer timescale corresponds to heating of the copper PCB, mounting bracket and still plate.

\subsection{Drift of optical modes}

\begin{figure} [htbp]
\centering
\includegraphics[width=\columnwidth]{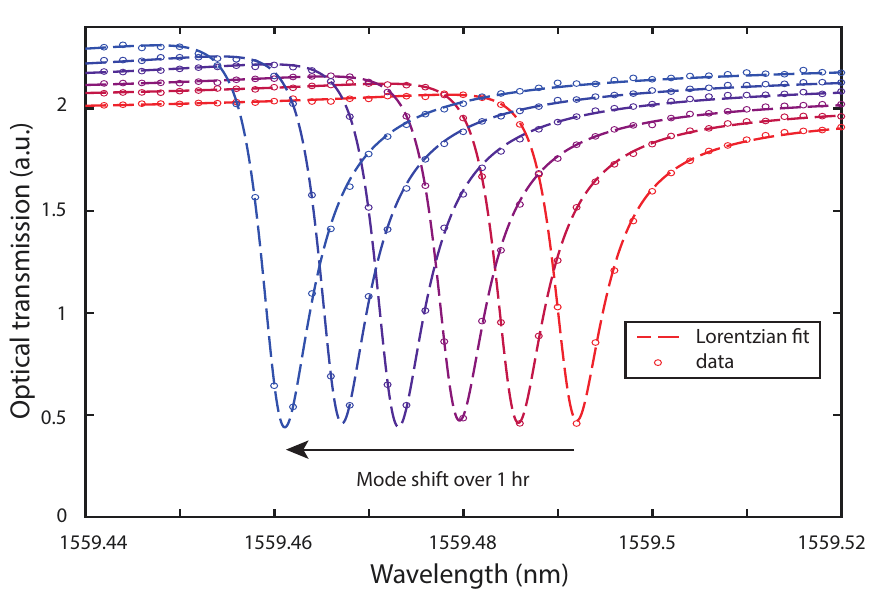}
\caption{The blue-shift of an optical mode over time.  The tunable laser was slowly stepped across the mode, from short to long wavelengths, with each point in the scan taking approximately 2 seconds (3 minutes per scan).  This was repeated for one hour and for clarity only every fourth scan is shown.  These repeated scans show that the mode drifted by about 30 pm over one hour, but during the experiment we also observed much faster, power dependent drift when trying to lock the laser to the optical mode.  For the scans pictured here the optical power to the device was 12 dBm.}
\label{fig:mode_drift}
\end{figure}

Another major effect which limited the conversion efficiency of the present device was frequency drift of the optical modes.  The modes drifted when the laser was tuned on-resonance, with the drift occurring more quickly for higher optical pump powers.  When no bias voltage is applied, the optical modes consistently shift to shorter wavelengths (blue-shift).  This shift towards shorter wavelengths is what would be expected from the photorefractive effect in lithium niobate \cite{Jiang2017,Liang2017}, while thermo-optic effects would cause the modes to shift to longer wavelengths (red-shift).  The steady blue-shift we observe is illustrated in Fig. \ref{fig:mode_drift}.  To keep the optical modes to be stable enough to perform the EO conversion measurement, we were limited to optical pump powers of 1 mW or less.

We also observe that when a bias voltage is applied to tune the wavelength of the modes, the optical modes tend to slowly shift in the opposite direction, back towards their original wavelength.  We attribute this to a more general photoconduction effect, where optically excited carriers in the LN arrange themselves to screen the applied electric field from the electrodes.  We note that the LN thin film in this work was congruently grown without any doping; doping with magnesium may be a potential avenue for reducing photorefraction \cite{Kip1998}. A more detailed analysis of optical drift in LN resonators will be presented in a future publication.

\end{document}